\documentclass[journal = ancac3]{achemso}
\setkeys{acs}{usetitle = true}

\usepackage[version=3]{mhchem} 

\author{Jariyanee Prasongkit}
\affiliation[Uppsala University]
{Division of Materials Theory, Department of Physics and Astronomy, Box 516, Uppsala University, SE-751 20 Uppsala, Sweden}
\alsoaffiliation[Nakhon Phanom University]
{Division of Science, Faculty of Liberal Arts and Science, Nakhon Phanom University, Nakhon Phanom, 48000, Thailand}
\author{Anton Grigoriev}
\author{Biswarup Pathak}
\affiliation[Uppsala University]
{Division of Materials Theory, Department of Physics and Astronomy, Box 516, Uppsala University, SE-751 20 Uppsala, Sweden}
\author{Rajeev Ahuja}
\affiliation[Uppsala University]
{Division of Materials Theory, Department of Physics and Astronomy, Box 516, Uppsala University, SE-751 20 Uppsala, Sweden}
\alsoaffiliation[Royal Institute of Technology]
{Applied Materials Physics, Department of Materials Science and Engineering, Royal Institute of Technology (KTH), SE-100~44 Stockholm, Sweden}
\author{Ralph H. Scheicher}
\email{ralph.scheicher@fysik.uu.se}
\affiliation[Uppsala University]
{Division of Materials Theory, Department of Physics and Astronomy, Box 516, Uppsala University, SE-751 20 Uppsala, Sweden}

\title[\texttt{achemso} demonstration]
{Transverse Electronic Transport through DNA Nucleotides with Functionalized Graphene Electrodes}

\begin{document}

\begin{abstract}
Graphene nanogaps and nanopores show potential for the purpose of electrical DNA sequencing, in particular because single-base resolution appears to be readily achievable. Here, we evaluated from first principles the advantages of a nanogap setup with functionalized graphene edges. To this end, we employed density functional theory and the non-equilibrium Green's function method to investigate the transverse conductance properties of the four nucleotides occurring in DNA when located between the opposing functionalized graphene electrodes. In particular, we determined the electrical tunneling current variation as a function of the applied bias and the associated differential conductance at a voltage which appears suitable to distinguish between the four nucleotides. Intriguingly, we observe for one of the nucleotides a negative differential resistance effect.
\end{abstract}

\maketitle


Despite existing methods for whole-genome analysis, there is a push towards the development of so-called third-generation sequencing techniques \cite{Schadt2010}, which would allow to operate on single molecules without amplification, and enable orders-of-magnitude longer base-read-lengths. Electrophoretic translocation of single-stranded DNA (ssDNA) through a nanopore is a potential avenue towards achieving this goal \cite{Chen2012,Bashir2011,Fyta2011,Branton2008}. While it was initially envisioned that the monitoring of ionic blockade currents could lead to the determination of the target DNA sequence, the focus has in the meanwhile shifted more towards transverse conductance measurements via nanopore-embedded electrodes \cite{Zwolak2005,Lagerqvist2006,Edel2010,Kaun2011,Sanvito2012} or via a combination of nanopores with nanowire field-effect transistors \cite{Lieber2012}. Due to the very small distance between adjacent bases in ssDNA, it is considered extremely challenging to fabricate embedded electrodes of sufficient sharpness that would couple electrically to only one base at a time. This problem of achieving single-nucleobase resolution is one of the main obstacles standing in the way of nanopore-based DNA sequencing, the other major challenge being control over the translocation process \cite{YuhuiHe2011a,YuhuiHe2011b}.

The proposal \cite{Postma:2010} to use atomically-thin graphene \cite{Neto:2009} created new momentum in the field, as it carries the potential for single-nucleobase resolution by avoiding the common size-mismatch between electrodes and target molecule \cite{Lofwander2011}. Subsequently, it was experimentally demonstrated by three research groups \cite{Dekker2010,Drndic2010,Golovchenko2010} that it is possible to translocate DNA through nanopores prepared in graphene, and the associated process was analyzed at the atomic level through molecular dynamics simulations \cite{Schulten:2011}. Furthermore, the translocation of DNA through nanopores fabricated in multilayered graphene-Al$_2$O$_3$ systems \cite{Bashir2012} could allow for fascinating new possibilities in detection and sequencing of DNA.

The principal capability of graphene nanopores \cite{Drndic:2008,Zettl:2009} and nanogaps \cite{vdZant:2011} to distinguish the four types of nucleobases has been evaluated from first principles \cite{Nelson:2010er,Prasongkit:2011gm,He2011a,Nikolic2012}. In these studies, the dangling bonds at the graphene edges were typically saturated with hydrogen \cite{Nelson:2010er,Prasongkit:2011gm,He2011a}, or, at least in one case, also with nitrogen \cite{Nikolic2012}. One could however also imagine other, more suitable functionalizing groups to be attached to the graphene edges (\emph{cf.} caption of Figure 6b in Ref.\ \citenum{Bashir2011}) which could lead to a better electronic coupling with the translocating ssDNA via the temporary formation of hydrogen bonds. Such functionalizing probes could be attached to only one electrode, coupling to the nucleobase part of DNA \cite{Lindsay2007,He:2008ct}, or to both electrodes, grabbing the phosphate group as well \cite{He:2009if,Pathak:2012}. The present study aims to explore the applicability of such functionalized graphene electrodes for the purpose of DNA sequencing or for detection of individual nucleotide molecules.

\begin{figure}[tp]
\begin{minipage}[ht]{1.0\linewidth}
\begin{center}
\includegraphics[scale =1]{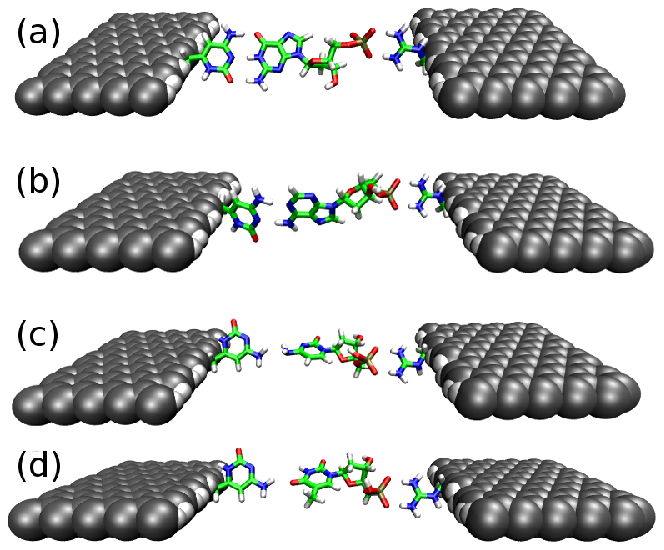}
\end{center}
\end{minipage}
\caption{Illustration of the double-functionalized graphene nano-electrodes for measuring the conductance of the four target nucleotides: (a) dGMP, (b) dAMP, (c) dCMP, and (d) dTMP. The graphene electrodes are functionalized by a phosphate-group-grabbing guanidinium ion on the right side and a reader-nucleotide (in the form of cytosine) on the left side.}\label{systems}
\end{figure}

In this work, we study the electronic transport properties of the four nucleotides deoxyadenosine monophosphate (dAMP), deoxythymidine monophosphate (dTMP), deoxyguanosine monophosphate (dGMP), and deoxycytidine monophosphate (dCMP), when located inside a functionalized graphene nanogap. Specifically, the armchair-edge of graphene is functionalized by a phosphate-group-grabbing guanidinium ion on one side (in the following referred to as the right-hand side) and a reader nucleotide on the other side (in the following referred to as the left-hand side). Here, we have considered cytosine as the reader nucleotide since this allows for the formation of the maximum number (\emph{i.e.}, three) of hydrogen bonds in the case of guanine-cytosine, the most stable base pair. In an earlier study, considering nanopore-embedded functionalized gold electrodes, we had tested all four nucleotides as probes and found cytosine to yield the best results in terms of target base differentiation for the purpose of DNA sequencing \cite{He:2008ct}. Functionalization of gold electrodes with cytosine was also independently suggested and investigated by another team of researchers \cite{Lindsay2007}. Also, guanine-functionalized electrodes have shown some promising results \cite{Hao2010}. As the ssDNA molecule is pulled through the nanogap by an electric field, the guanidinium will grab each target nucleotide passing between the scanning setup by forming temporary hydrogen bonds with the phosphate groups and, simultaneously, the reader nucleotide also forms hydrogen bonds with the base part of the target nucleotide. The illustration of the two-probe system, consisting of the target nucleotide placed between functionalized graphene electrodes, is presented in~\ref{systems}. This setup can not only help to improve the electronic coupling between electrodes and nucleotides, but it can also lead to a preferred orientation of the target nucleotides relative to the electrodes, thus reducing noise, and furthermore slow down the translocation speed of the DNA, allowing more time for each individual measuring process. We emphasize that the advantageous aspects of double-functionalization, albeit not for graphene edges, but rather for the case of gold electrodes has been investigated by both He \emph{et al.} \cite{He:2009if} and us \cite{Pathak:2012}.


The geometrical structure of the cytosine-target nucleotide-guanidinium part was first relaxed by using the Gaussian09 package \cite{Gaussian09} with B3LYP/6-31G$^*$ and then placed in the gap between hydrogen-terminated armchair graphene edges. The graphene electrodes are separated from another by 23.82 \AA\ (measured from H to H), which is maintained throughout the calculations. The central region includes 5.05 \AA\ of the left and right parts of the graphene electrodes to screen any perturbation effects caused by the nucleotide and functionalized molecule. The large vacuum distance ($\sim$10 \AA) between the nucleotides is sufficient to avoid the interaction between repeated images due to the periodic boundary condition along the $x$-direction. Each two-probe system contains around 230 atoms.

The whole system was optimized by using density functional theory (DFT) as implemented in the SIESTA package \cite{Soler2002}. The generalized gradient approximation (GGA) \cite{Lee:1988} was employed to approximate the exchange and correlation functional. For all elements, we used basis sets including polarization orbitals (SZP), which is good enough for the nucleotide-graphene systems. The atomic core electrons are modeled with Troullier-Martins norm-conserving pseudopotentials \cite{Troullier1991}. Only the $\Gamma$-point was considered for Brillouin zone sampling owing to the large cell size. The mesh cutoff value is 170 Ry for the real-space grid.

The transport properties were then carried out with the quantum transport code TranSIESTA \cite{Brandbyge:2002ck}, combining DFT and the non-equilibrium Green's function (NEGF) technique. Since TranSIESTA uses SIESTA as its DFT platform, the basis sets and the real-space grid employed in the transport calculations are identical to those described above for the geometrical optimization part. Within the NEGF approach, graphene electrodes are extended to infinity (${z\to\pm\infty}$), meaning we have semi-infinite electrodes to either side. When bias voltage is applied, the current is driven to flow through the system, obtained from integration of the transmission spectrum:
\begin{equation}
I(V_{b}) =
\dfrac{2e}{h}\int_{\mu_{R}}^{\mu_{L}}T(E,V_{b})[f(E-\mu_{L})-f(E-\mu_{R})]
dE, \label{equation1}
\end{equation}
where $T(E,V_b)$ is the transmission probability of electrons
incident at an energy $E$ from the left to the right electrode under
an applied bias voltage $V_b$, and $f(E-\mu_{L,R})$ is the
Fermi-Dirac distribution of electrons in the left (L) and right (R)
electrode with the respective chemical potential $\mu_{L} =
E_{\mathrm{F}}+V_b/2$ and $\mu_{R} = E_{\mathrm{F}}-V_b/2$ shifted
respectively up or down relative to the Fermi energy
$E_{\mathrm{F}}$.

\begin{figure*}[tp]
\begin{minipage}[ht]{1.0\linewidth}
\begin{center}
\includegraphics[width = 14 cm]{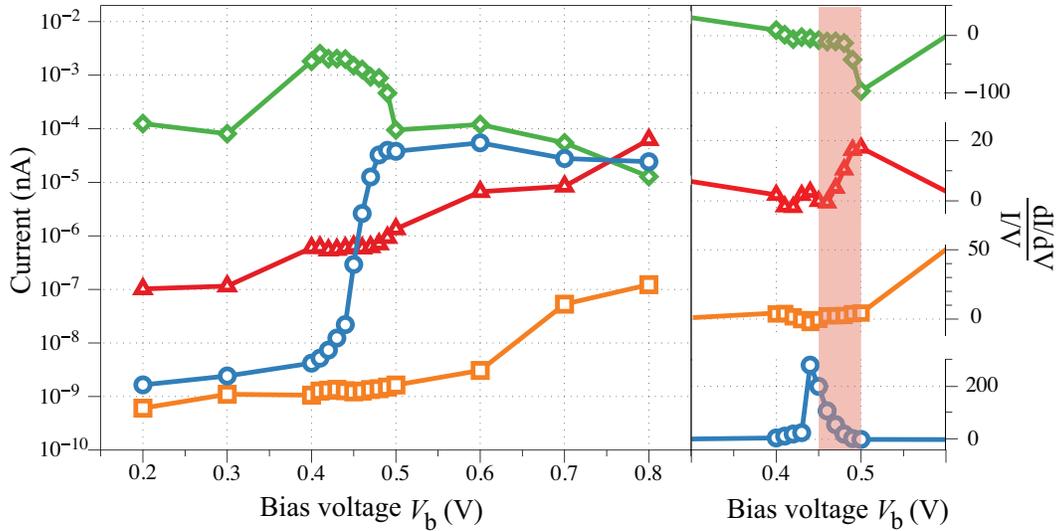}
\end{center}
\end{minipage}
\caption{Current-voltage curves plotted on a semi-logarithmic scale in the left panel for the four target nucleotides: dGMP (green diamonds), dAMP (red triangles), dTMP (blue circles), and dCMP (orange squares). In the right panel, the normalized first derivative of the tunneling current, $(dI/dV)/(I/V)$, is plotted, showing a possibility for distinguishing the four nucleotides from a single voltage scan, as discussed in the text.}\label{iv}
\end{figure*}

We first discuss the \textit{I-V} characteristics (IVC) of our studied system described above. As it can be seen in the left panel of \ref{iv}, dGMP shows exceptionally high current throughout the entire bias voltage ($V_b$) range. The difference between dGMP and dAMP is about 3 orders of magnitude for 0 < $V_b$ < 0.4 V. In addition, the difference in the current of both dGMP and dAMP to dTMP and dCMP is more than 1 order of magnitude. Therefore, for the low bias range ($V_b$ < 0.4 V), dGMP and dAMP might be identifiable based on their current signature. However, dCMP and dTMP appear indistinguishable from each other.

The pronounced distinction between dCMP and dTMP starts for $V_b$ > 0.4 V, when the current of dTMP rises rapidly above that of dCMP by almost 5 orders of magnitude, before plateauing at $V_b$ = 0.5 V (\ref{iv}). In the bias range $0.5 \leq V_b \leq 0.8$ V, the difference between the current of dTMP and dCMP gradually decrease to eventually 2 orders of magnitude due to the steady rise of current for dCMP. Nonetheless, one can see from \ref{iv} that for the purpose of distinguishing between dTMP and dCMP, one should look at $V_b$ > 0.5 V, where a low current would indicate the presence of dCMP, whereas a high current would lead to the positive identification of dTMP.

We have thus shown a principal capability for sequencing nucleotides by considering the IVC, in which two measurements at two different bias voltages are needed for proper base distinction: one below 0.4 V to distinguish between dGMP, dAMP and either dTMP or dCMP, and another one above 0.5 V to resolve the remaining ambiguity between dTMP and dCMP. In an effort to simplify the detection process and avoid the need for measurement at two different bias voltages, we have explored the possibility of distinguishing nucleotides by differential conductance, \emph{i.e.}, the first derivative of the current with respect to the bias voltage, $dI/dV$. This approach possesses several advantages over a mere current measurement, as we have discussed previously \cite{He2010}. For the calculation of the differential conductance, we used steps of 0.01 V for $V_b$ between 0.4 and 0.5 V. As it can be seen from the right panel of \ref{iv}, dTMP can be distinguished at about 0.45 V by an exceptionally high $(dI/dV)/(I/V)$ value. Also, dGMP can be uniquely identified at 0.50 V by the unusual negative value of $dI/dV$. The other two nucleotides show either moderate $dI/dV$ values (dAMP) or virtually zero $dI/dV$ (dCMP), leading to their respective distinction.

Now we turn to a more detailed discussion of how the current behaves under an applied voltage for each nucleotide. As seen in the left panel of \ref{iv}, at a bias voltage below 0.4 V, the respective current magnitudes appear to yield the ordering dGMP > dAMP > dTMP > dCMP. When the voltage rises above 0.4 V, dGMP exhibits negative differential resistance (NDR), with the current dropping by more than 2 orders of magnitude within the bias interval 0.4 V $\leq V_b \leq$ 0.8 V. Moreover, the current associated with dTMP tremendously increases by 4 orders of magnitude in the 0.4 V $\leq V_b \leq$ 0.5 V interval. For dAMP and dCMP, the current magnitudes are seen to increase in stages, but mostly so for $V_b$ > 0.6 V, especially in the case of dCMP.

\begin{figure*}[tp]
\begin{minipage}[ht]{1.0\linewidth}
\begin{center}
\includegraphics[width = 16 cm]{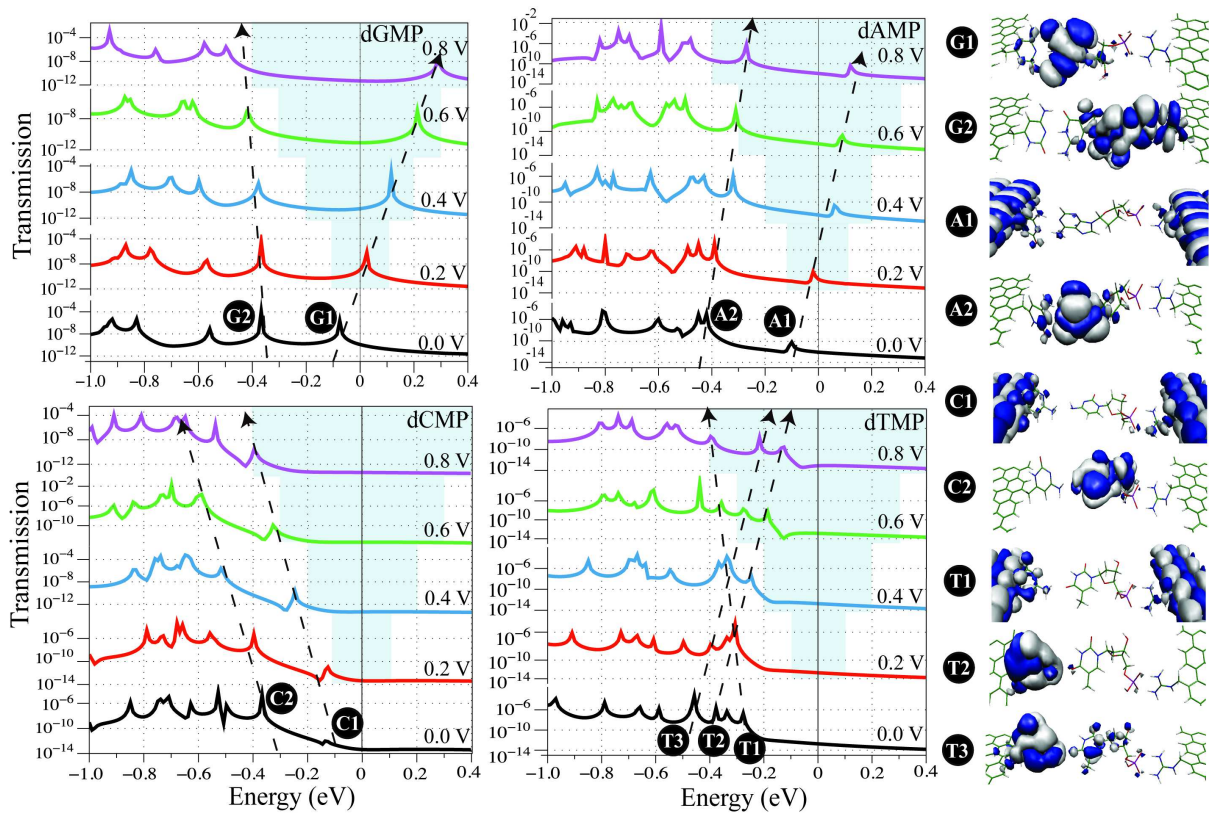}
\end{center}
\end{minipage}
\caption{The bias-dependent transmission of the four target nucleotides, dGMP (top left panel), dAMP (top right panel), dCMP (bottom left panel), and dTMP (bottom right panel), as a function of energy \textit{E}. The shaded area indicates the growing bias voltage window. The rightmost panel shows isosurface plots of the molecular orbitals responsible for those transmission peaks labeled by the letters G, A, C, and T, and respective numbers in the four panels to the left.}\label{transmission}
\end{figure*}
\begin{figure}[tp]
\begin{minipage}[ht]{1.0\linewidth}
\begin{center}
\includegraphics[width = 8 cm]{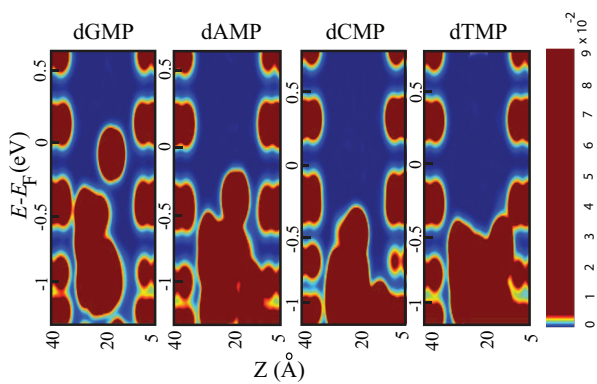}
\end{center}
\end{minipage}
\caption{The zero-bias DOS of the target nucleotides (a) dGMP, (b) dAMP, (c) dCMP, (d) dTMP, when connected to the functionalized graphene electrodes. Zero energy is aligned with the Fermi level. Red and blue colors correspond to high and low density of states, respectively. The scale is chosen to emphasize the localization of the molecular states in the gap between the graphene electrodes.}\label{dos}
\end{figure}

To achieve a better understanding of these IVC features, we analyze the bias-dependent transmission spectra (\ref{transmission}) together with the DOS at zero-bias (\ref{dos}) of all four target nucleotides when located between the functionalized graphene electrodes. We find that the motion of transmission resonance peaks under applied bias depends on the electronic structure of each nucleotide and on the coupling of the molecular states of the target nucleotides to the functionalized graphene edges. Below we describe and discuss this behavior in detail for each of the four nucleotides. The energy levels in the cytosine functional group rise following the chemical potential of the left electrode as the bias voltage is increased, and are lowered in the guanidinium following the chemical potential of the right electrode.

\textit{dGMP}: From \ref{dos} it can be seen that the DOS peak at zero bias corresponds well to the positions of transmission peaks and the molecular orbitals shown in \ref{transmission} at zero bias voltage. When compared to other nucleotides, the highest occupied molecular orbital (HOMO) state of dGMP localized on the base part is closest to the Fermi energy. As a result, dGMP yields the highest magnitude of current in proximity to zero bias, \emph{i.e.}, in the low bias range.

\ref{transmission} shows that the HOMO, labeled as G1, moves upward in energy following the rising chemical potential of the left electrode, while HOMO--1 (G2) moves downward in energy following the lowering chemical potential of the right electrode. In the right panel of \ref{transmission}, the isosurface plots of the molecular orbitals corresponding to the HOMO and HOMO--1 are drawn. The HOMO is well localized in the guanine base and extends into some parts of the functionalizing cytosine probe forming hydrogen bonds, while the HOMO--1 is localized in the part of the sugar-phosphate and guanidinium ion. Thus, the HOMO state follows the right electrode, whereas the HOMO--1 state follows the left electrode over the entire bias region. As a consequence, with increasing bias voltage, the HOMO is shifted towards higher energy and stays within the bias window for the whole bias range, while the HOMO--1 is shifted downward in energy and remains outside the bias window.

\begin{figure}[tp]
\begin{minipage}[ht]{1.0\linewidth}
\begin{center}
\includegraphics[width = 8 cm]{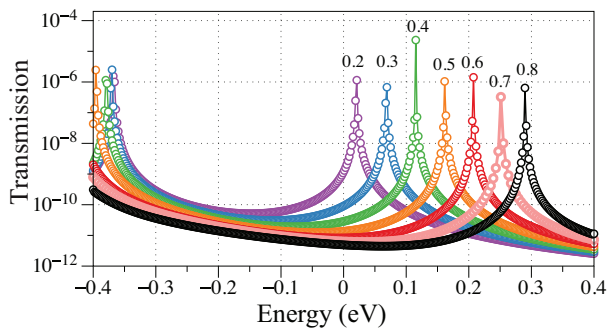}
\end{center}
\end{minipage}
\caption{A comparison of the transmission spectra for dGMP calculated at different bias voltages ranging from 0.2 V to 0.8 V with bias steps of 0.1 V. }\label{baseline}
\end{figure}

The main transport channel of dGMP is the HOMO peak which lies in the bias window for $V_b$ < 1 V, as seen in \ref{transmission}. With increasing bias voltage, the HOMO peak shifts uniformly to higher energies since the HOMO state is pinned to the upward-in-energy-moving electronic states of the left electrode. As the bias increases, a comparison of transmission spectra calculated at the bias voltage from 0.2 V to 0.8 V (\ref{baseline}) reveals a sharp rise in the transmission resonance peak at $V_b$ = 0.4 V, resulting in an increase of the current by about one order of magnitude. When the applied voltage is raised to 0.5 V, the height of the transmission peak drops, the current decreases, and the NDR occurs. At 0.5 V $\leq V_b \leq$ 0.8 V, the HOMO peak continues to shift towards higher energy while the transmission baselines in the region between the HOMO and the HOMO--1 peak are moving down, and therefore the current gradually drops.

NDR is of great interest for potential applications in molecular electronics devices. Different proposed mechanisms have been employed to explain the NDR \cite{Chen:1999,Lyo:1989,Pati:2008bj}. However, since the focus of the present work is on DNA sequencing applications, we will return to the NDR effect in dGMP further below in the manuscript, after the discussion of the other three nucleotides has been concluded.

\textit{dAMP}: From \ref{transmission} it can be seen that there are two transmission peaks (A1 and A2) which contribute to the current in the entire bias interval examined by us (\emph{i.e.}, 0 V $\leq V_b \leq$ 0.8 V). However, only the A1 peak behaves as a resonant channel in the voltage window at low bias with the molecular orbital of the A1 peak fully delocalized over both graphene electrodes. The A2 peak is localized on the adenine base part and expanded to some parts of the cytosine probe. The positions of the resonance peaks in the transmission agree well to the alignment of the DOS peaks (\ref{dos}). Note that the molecular orbital of the A2 peak corresponds to the HOMO of the isolated adenine base. Hybridization of the adenine HOMO state with the states in the graphene electrodes through the cytosine probe leads to the A2 broadening. For $V_b$ < 0.6 V, the current has not been significantly affected by the A1 peak lying in the bias window; a rise in the current at that range results from a shoulder of the HOMO peak as it comes into the bias window. With increasing bias voltage, the HOMO peak moves upward in energy following the upward-in-energy-moving states of the left electrode, since the state is well localized in a region controlled by the left lead. At $V_b$ > 0.6 V, the HOMO enters the bias window resulting in a monotonic increase in the current.

\textit{dTMP}: The finite, yet low current of dTMP at $V_b < 0.4$ V (\ref{iv}) results from the alignment of the molecular levels relative to the Fermi energy. As seen from \ref{transmission}, there is no resonance peak for $V_b$ < 0.4 V. The molecular orbital (see right panel of \ref{transmission}) corresponding to T1 is fully delocalized over both graphene electrodes, while the T2 and T3 peaks are mainly localized on the cytosine probe attached to the left electrode. As a result, the T2 and T3 peaks shift upward in energy following the upgoing left electrode. The T1 peak shifts downward in energy following the downgoing right electrode because there is no strong coupling to the left electrode. With increasing bias, the T1, T2, and T3 peaks all end up eventually within the bias window, resulting in a sudden rise in current at $V_b$ > 0.4 V.

From the DOS of dTMP shown in \ref{dos}, one can observe that the states are delocalized throughout the junction. Consequently, once the applied bias is sufficiently large (\emph{i.e.}, above 0.4 V) for the states to enter into the bias window, electrons can go through resonant tunneling channels from one electrode to the other, and the current abruptly increases as a result (\ref{iv}).

\textit{dCMP}: The DOS plot in \ref{dos} reveals that dCMP does not couple well with the left electrode, \emph{i.e.}, the cytosine probe cannot interact well with its ``twin'' nucleotide. In addition, the C1 and C2 peaks, which are the closest ones to the Fermi energy, shift away when voltage is applied (\ref{transmission}). At $V_b = 0.7$ V, the C1 peak starts to come into the bias window, resulting in an increased current (\ref{iv}). However, the C1 peak is not a dominant resonance channel, since the molecular orbital corresponding to that peak is mainly localized on the electrode (see right panel of \ref{transmission}). Compared to the other three nucleotides, the current values of dCMP are therefore the lowest over the entire bias region.

\begin{figure}[tp]
\begin{minipage}[ht]{1.0\linewidth}
\begin{center}
\includegraphics[width = 16 cm]{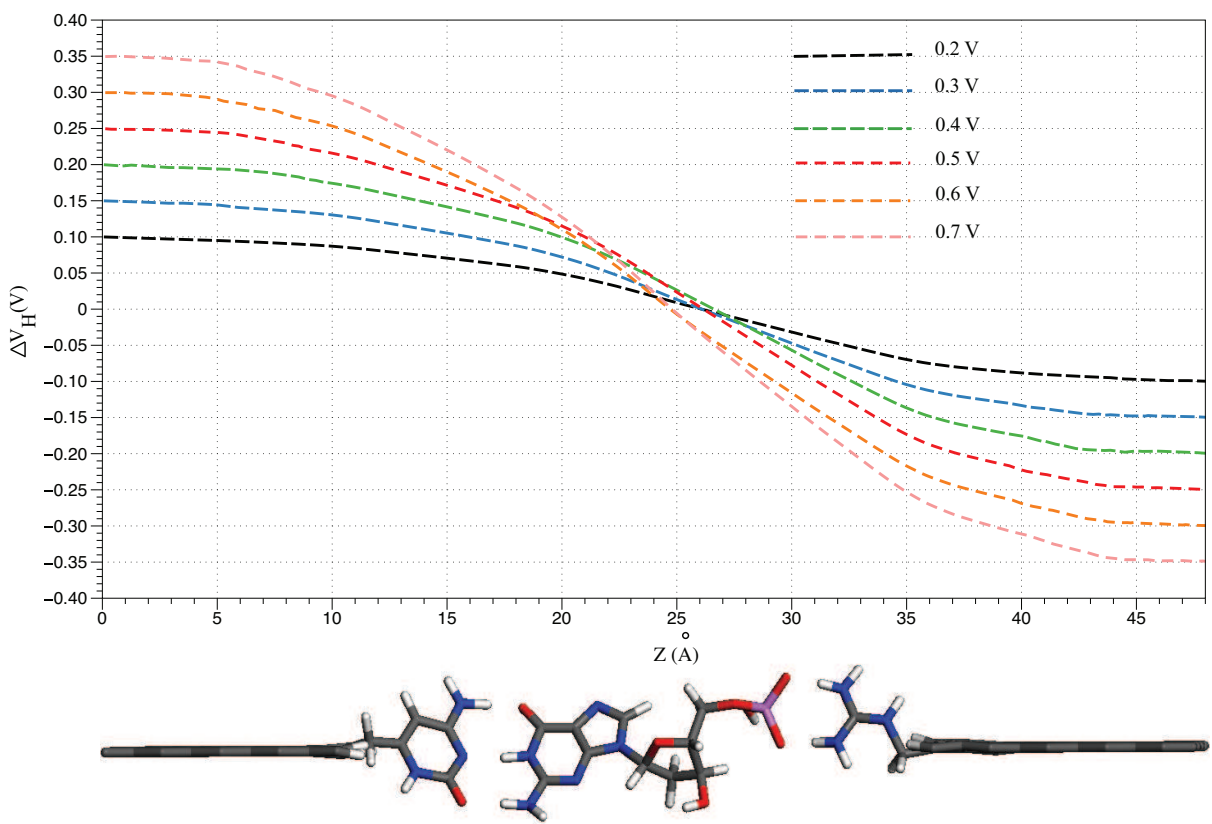}
\end{center}
\end{minipage}
\caption{Electrostatic potential showing the voltage drop across a double-functionalized graphene junction with dGMP located in between the electrodes.}\label{pot}
\end{figure}

We now turn to discuss the computationally observed NDR effects in dGMP. Typically, NDR has been found for semiconducting electrodes in various molecular devices due to the misalignment of electronic states between the electrodes and the molecules \cite{Ribeiro:2008in,Rakshit:2004db}. However, graphene sheets with armchair edges possess no band gap. For our system, the occurrence of NDR is due to the asymmetric position of the HOMO of dGMP (\ref{dos}). The HOMO of dGMP is localized on the base, close to the left lead. Under an applied bias, electrons from the HOMO state will tunnel to the right electrode. However, with increasing bias voltage the coupling between the target nucleotide and the right electrode is weakening. This is manifested in the overall decrease of the transmission curve with bias (\ref{transmission}). Such a conclusion is supported by an analysis of the potential drop of the system (\ref{pot}), where we observe the main potential drop taking place at a region of sugar and phosphate groups, and extends on the base part at high bias.

The NDR effect resulting from asymmetric molecular junctions has been suggested recently \cite{Leijnse:2011ce}. Furthermore, the NDR behavior in this model might be similar to that found in a scanning tunneling microscope (STM), where the tunneling of electrons through a potential barrier results in a decrease in current at finite bias \cite{Bevan:2008ey,Chen:2007hc}.

\begin{figure}[tp]
\begin{minipage}[ht]{1.0\linewidth}
\begin{center}
\includegraphics[width = 12 cm]{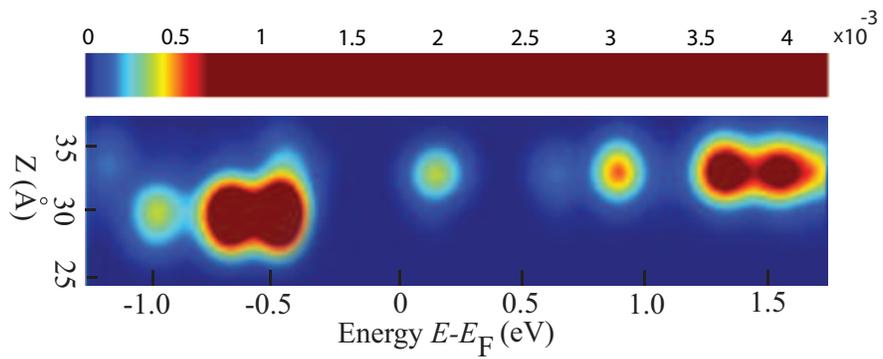}
\end{center}
\end{minipage}
\caption{The zero-bias DOS of the guanidinium ion connected to the right-hand graphene electrode. Zero energy is aligned with the Fermi level. Red and blue colors correspond to high and low density of states, respectively.}\label{func}
\end{figure}

At $V_b$ = 0.4 V, we observe a sharp rise in the transmission resonance peak. We demonstrate the discreteness in the molecular energy level of the guanidinium ion (\ref{func}). As a result, electrons from the HOMO can tunnel to the state of the guanidinium ion (close to the Fermi energy) at about 0.4 V, and the corresponding transmission peaks become higher, leading to a significant increase in the current. When the bias voltage changes, a mismatch of state alignment results in the seen current drop. Hence, we propose that the NDR behavior of dGMP in our setup arises from (i) the asymmetric HOMO of dGMP and (ii) an existence of discrete guanidinium states close to the Fermi level which can be utilized by electrons to tunnel into.


In conclusion, it appears from our study that double-functionalized graphene electrodes might offer some advantages over merely hydrogenated graphene edges for the purpose of DNA sequencing. The two functionalizing groups studied by us (namely, cytosine and guanidinium ion) have been respectively designed to optimally couple to the different parts of ssDNA (namely, the nucleobase and the phosphate group). Through the formation of temporary H-bonds, the electronic coupling could be enhanced and the nucleotide would be stabilized between the opposing graphene electrodes, potentially allowing for less orientational fluctuations and thus supposedly reducing the associated variance in the tunneling current signals. Furthermore, the functionalizing groups should also help to slow down the translocation process, providing more time for each individual measurement. Compared to non-functionalized graphene electrodes \cite{Prasongkit:2011gm}, we find here for the functionalized system that the shift of molecular orbitals closer towards the Fermi level allows for better electric recognition of the nucleobase identity, and that potentially a single measurement of the differential conductance at a bias voltage of around 0.45 to 0.50 V could lead to an unambiguous distinction between all four nucleotides.

Of course, it remains to be seen whether these predicted advantages would actually materialize in any experimental setup. Functionalization of nano-electrodes undoubtedly adds another layer of complexity to any fabrication process (although it could be argued that some functionalizing groups might be introduced to graphene edges as a convenient side-effect of the nanopore creation). The realization of the above listed benefits of functionalization depends however on whether or not the passing ssDNA can ``see'' the functionalizing probe molecules. For that to happen, the pore diameter must not be too large and the translocation speed cannot exceed a certain limit. Molecular dynamics simulations are a very helpful computational tool in this respect to explore promising pore designs that would lead to an actual utilization of functionalized electrodes, and such investigations (which are beyond the scope of the present study) are urged to be carried out on this system.

Finally, we note that for the target nucleotide deoxyguanosine monophosphate (but not for the other three nucleotides), we found a negative differential resistance effect in our calculated $I-V$ curve, which was explained by us as a combined consequence of the asymmetric HOMO of the deoxyguanosine monophosphate and certain guanidinium states near the Fermi level through which electrons can tunnel.

\section*{Acknowledgements}

Financial support is gratefully acknowledged from the Royal Thai Government, the Carl Tryggers Foundation, the Uppsala University UniMolecular Electronics Center (U$^3$MEC), Wenner-Gren Foundations, the Swedish Foundation for International Cooperation in Research and Higher Education (STINT), and the Swedish Research Council (VR, Grant No. 621-2009-3628). The Swedish National Infrastructure for Computing (SNIC) and the Uppsala Multidisciplinary Center for Advanced Computational Science (UPPMAX) provided computing time for this project.



\newpage

\end{document}